\documentclass[amsmath,superscriptaddress,showpacs,pre]{revtex4}
\usepackage{graphicx}
\usepackage{amsmath}

\usepackage{epsfig}
\usepackage{graphics}
\usepackage{latexsym}
\usepackage{amsfonts}
\usepackage{amssymb}
   
\usepackage{plain}

\begin{document}

\title{Hysteresis and complexity in the zero-temperature mean-field RFIM: the soft-spin version}

\author{M. L. Rosinberg}
\email{mlr@lptmc.jussieu.fr}
\affiliation{ Laboratoire de Physique Th\'eorique de la Mati\`ere Condens\'ee, CNRS-UMR 7600, Universit\'e Pierre et Marie Curie, 4 place Jussieu, 75252 Paris Cedex 05, France}

\author{T. Munakata}
\email{munakata@amp.i.kyoto-u.ac.jp}
\affiliation{Department of Applied Mathematics and Physics, Graduate School of Informatics, Kyoto University, Kyoto 606-8501, Japan}

\begin{abstract}
We study the energy landscape of the soft-spin random field model in the mean-field limit and compute analytically the quenched complexity of the metastable states as a function of their magnetization and energy at a given external magnetic field. The shape of the domain within which the complexity is positive (and the number of typical metastable states grows exponentially with system size) changes with the amount of disorder and becomes non-convex and disconnected at low disorder. As a consequence, phase transitions occur both at equilibrium and out of equilibrium along the saturation hysteresis loop.  We focus on the zero complexity curve in the field-magnetization plane and its relationship with the hysteresis loop. We also study the response of the system when the magnetization is externally controlled instead of the magnetic field. The main features of the model that should survive in finite dimensions are discussed.
\end{abstract}

\pacs{64.60.av, 75.60.Ej, 75.10.Nr}

\maketitle

\def\be{\begin{equation}}
\def\ee{\end{equation}}
\def\bea{\begin{align}}
\def\eea{\end{align}}

\section{Introduction}

The response of systems with quenched disorder to an applied field or force is typically hysteretic and noisy. A well-known example is the Barkhausen noise in ferromagnetic materials that results from the intermittent motion of magnetic domain walls in response to a change in the external magnetic field\cite{DZ2006}. This behavior, which is observed in a wide variety of physical systems, from superconductors\cite{FWNL1995} to martensitic shape-memory alloys\cite{VOMRP1994}, can be related to the complicated structure of the energy landscape, with a huge number of local minima (or metastable states) separated by large energy barriers, which makes thermal fluctuations irrelevant on experimental time scales. A jump or avalanche then corresponds to the  disappearance of a local minimum in the landscape as the external field is changed. Much effort has been devoted in recent years to describe the avalanche statistics in various models and to explain the origin of the power-law distributions that are ubiquitously observed. Depending on the model and the details of the dynamics, scale-invariance results from self-organized criticality or requires fine tuning of the system parameters\cite{KS2000}. 

In the theoretical description, one  usually considers a zero-temperature dynamical evolution,  which implies that the system, prepared in some known initial configuration, visits a deterministic sequence of metastable states (similarly, at the thermodynamic equilibrium, the system jumps from one global energy minimum to another while the field is changed). Therefore, the stochastic character of the output signal only reflects the intrinsinc disorder in the system and the whole information about the nonequilibrium response, for instance at criticality, is encoded in the energy landscape. This means that it should be possible in principle (but maybe nontrivial in practice) to describe the disorder-averaged properties of the response using a statistical description of the metastable states, that is {\it without} taking into account the entire history of the system. Of course, this supposes that the nonequilibrium path can be characterized unambiguously, for instance by some ``extremal'' property. This is illustrated in the present work where we study a simple mean-field model for which a complete analytical description of the distribution of the metastable states can be reached.  This model was introduced in Ref.\cite{DS1996} as the starting point for a renormalization-group description of the nonequilibrium, zero-temperature random-field Ising model (RFIM), and it may be also the basis for a similar study of the energy landscape.

The zero-temperature Gaussian RFIM with a metastable dynamics is a prototype for a large class of disordered systems with avalanchelike behavior\cite{SDP2006}. The main feature in three and higher dimensions is the existence of a nonequilibrium critical point that separates two different regimes of avalanches\cite{S1993}. Above the critical disorder, all avalanches (i.e. Barkhausen jumps) are of microscopic size and the saturation hysteresis loop, obtained by cycling the magnetic field adiabatically from large negative to large positive values and back, is smooth in the thermodynamic limit. Below the critical disorder, there is a macroscopic avalanche at a certain field and the hysteresis loop is discontinuous. In the following, we consider a soft-spin version of the RFIM where a spin can take any value between $-\infty$ and $+\infty$, and we study the mean-field limit where every spin interacts equally with every other spin. The model can then be viewed as a collection of identical bistable (or Preisach) units interacting via a mean-field term. This type of model is very popular in the hysteresis community (see e.g. Ref.\cite{B1998}) and its properties have been studied in great detail. However, as far as we know, the issues that we want discuss here have not been addressed. 

The relationship between the hysteresis loop in the zero-temperature RFIM and the distribution of the metastable states in the field-magnetization plane was investigated analytically and numerically in a series of recent papers\cite{DRT2005,PRT2008,RTP2008a,RTP2008b}. The information about the distribution of the metastable states in the $H-m$ plane is encoded in the ``quenched'' complexity $\Sigma_Q(m,H)$ which is the logarithm (divided by the number of spins) of the typical number of metastable states at the field $H$ with a given magnetization per spin $m$. The number of metastable states grows exponentially with the system size when $\Sigma_Q(m,H)$ is positive, and it appears that the curve $\Sigma_Q(m,H)=0$ exactly coincides with the saturation hysteresis loop in the large-disorder regime\cite{DRT2005,PRT2008,RTP2008b}. The situation at low disorder is more complicated but more interesting, and it has been conjectured that the boundary of the domain of existence of the metastable states coincides with the hysteresis loop obtained by externally controlling the magnetization instead of the magnetic field. This is an issue which is also relevant to some experimental situations\cite{B1998,BRIMPV2007}. Our aim in this work is to check  analytically this whole scenario in the mean-field model. We shall also consider energetics aspects which play an important role at low disorder where phase transitions occur, depending on the energy of the metastable states. This information is encoded in the magnetization and energy dependent complexity $\Sigma_Q(m,e,H)$. 

The outline of the paper is as follows. In Sec.  II, we present the model and briefly review the hysteretic behavior  discussed in Ref.\cite{DS1996}. In Sec. III, we compute the complexity, first as a function of magnetization only and then as a function of magnetization and energy. We focus on the low-disorder regime where phase transitions occur, both in equilibrium (i.e. in the ground state) and out of equilibrium along the hysteresis loop. In Sec. IV, we compute the nonequilibrium response obtained by controlling the magnetization and discuss its relationship with the usual hysteresis loop and the complexity. Section V concludes with a brief discussion.

\section{Model and hysteresis loop}

We consider a collection of $N$ soft spins interacting via the Hamitonian 
\begin{equation}
\label{Eq1}
 {\cal H} = -\frac{J}{2N}\sum_{i\ne j}s_is_j-\sum_i(H+h_i)s_i+\sum_i V(s_i)
\end{equation}
where $J>0$ is a ferromagnetic coupling, $H$ is an external uniform field, and $\{h_i\}$ is a set of quenched random fields drawn independently from a probability distribution ${\cal P}(h)$ (in practice, this will be a Gaussian distribution with zero mean and standard deviation $\Delta$). $V(s_i)$ is a double-well potential that mimics the two states of the hard-spin model. As in Ref.\cite{DS1996}, we will choose 
\begin{equation}
\label{Eq2}
V(s) = \left\{\begin{array}{ll} \frac{k}{2}(s+1)^2 &  \mbox{for $s<0$}\\
\frac{k}{2}(s-1)^2 & \mbox{for $s>0$} \ .
\end{array}\right. 
\end{equation}
with $k>J$ (this condition ensures that the magnetization remains finite at any field $H$\cite{DS1996}). The ``metastable'' states are the local minima of the Hamiltonian where each spin satisfies $\partial {\cal H}/\partial s_i=0$, i.e.
\begin{equation}
\label{Eq3}
s_i-\mbox{sign}(s_i) = \frac{H+Jm+h_i}{k} 
\end{equation}
where $m=(1/N) \sum_i s_i$ is the magnetization per spin. Clearly, this equation has no negative (resp. positive) solution for $h_i>-H-Jm+k$ (resp. $h<-H-Jm-k$). On the other hand, $s_i$ can be either positive or negative for $-H-Jm -k<h_i<-H-Jm+k$. As a consequence, there is  a field interval where the  number of metastable states with magnetization $m$ grows exponentially with $N$ (and the corresponding complexity is positive). This contrasts with the mean-field Ising model studied in Ref.\cite{RTP2008a} where this number is finite (and actually very small) in the thermodynamic limit.

At $T=0$, when increasing the external field adiabatically from $-\infty$, a spin remains negative as long as Eq. (\ref{Eq1}) admits a solution with $s_i<0$. When this is no more possible, the spin moves to the ``up''  potential well ($s_i>0$) and stays at the bottom of the new well. This move is equivalent to a spin flip in the hard-spin model. Therefore, for given field $H$ and given magnetization $m$, all spins with $h_i<-H-Jm+k$ are negative and all spins with $h_i>-H-Jm+k$ are positive. The self-consistency condition for the average magnetization then reads
\begin{align}
\label{Eq4}
m_\uparrow(H) &= \int_{-\infty}^{-H-Jm_\uparrow+k} dh {\cal P}(h) [(H+Jm_\uparrow+h)/k -1]+\int_{-H-Jm_\uparrow+k}^{\infty} dh {\cal P}(h) [(H+Jm_\uparrow+h)/k +1] \  ,
\end{align}
which yields\cite{DS1996}
\begin{equation}
\label{Eq5}
m_\uparrow(H)=\frac{H+k}{k-J}-\frac{2k}{k-J}\int_{-\infty}^{-H-Jm_\uparrow+k}dh {\cal P}(h) \ .
\end{equation}
 Similarly, when decreasing $H$ from $+\infty$, the magnetization is solution of the self-consistent equation
\begin{equation}
\label{Eq6}
m_\downarrow(H)=\frac{H+k}{k-J}-\frac{2k}{k-J}\int_{-\infty}^{-H-Jm_\downarrow-k}dh {\cal P}(h) \ .
\end{equation}

Depending on the disorder strength, these equations admit one or three solutions in a certain interval of $H$ and the resulting hysteresis loop is then smooth or discontinous in the thermodynamic limit (the magnetization can only vary monotonously with the field and the ``unstable'' solutions of Eqs. (\ref{Eq5}) or (\ref{Eq6}) that correspond to a negative susceptibility $dm/dH$ have no meaning in this context). Unlike the hard-spin mean-field model, hysteresis is always present. Specifically, for a Gaussian distribution ${\cal P}(h)$, the  loop is smooth for $\Delta>\Delta_c$ and discontinuous for $\Delta<\Delta_c$, with $\Delta_c/J=\sqrt{2/\pi}\  k/(k-J)$\cite{DS1996}. This is illustrated in Fig. 1. (Hereafter, without loss of generality, we shall always choose $k=3$ in the numerical calculations and take $J=1$ as the energy unit.)

\begin{figure}[hbt]
\includegraphics[width=10cm]{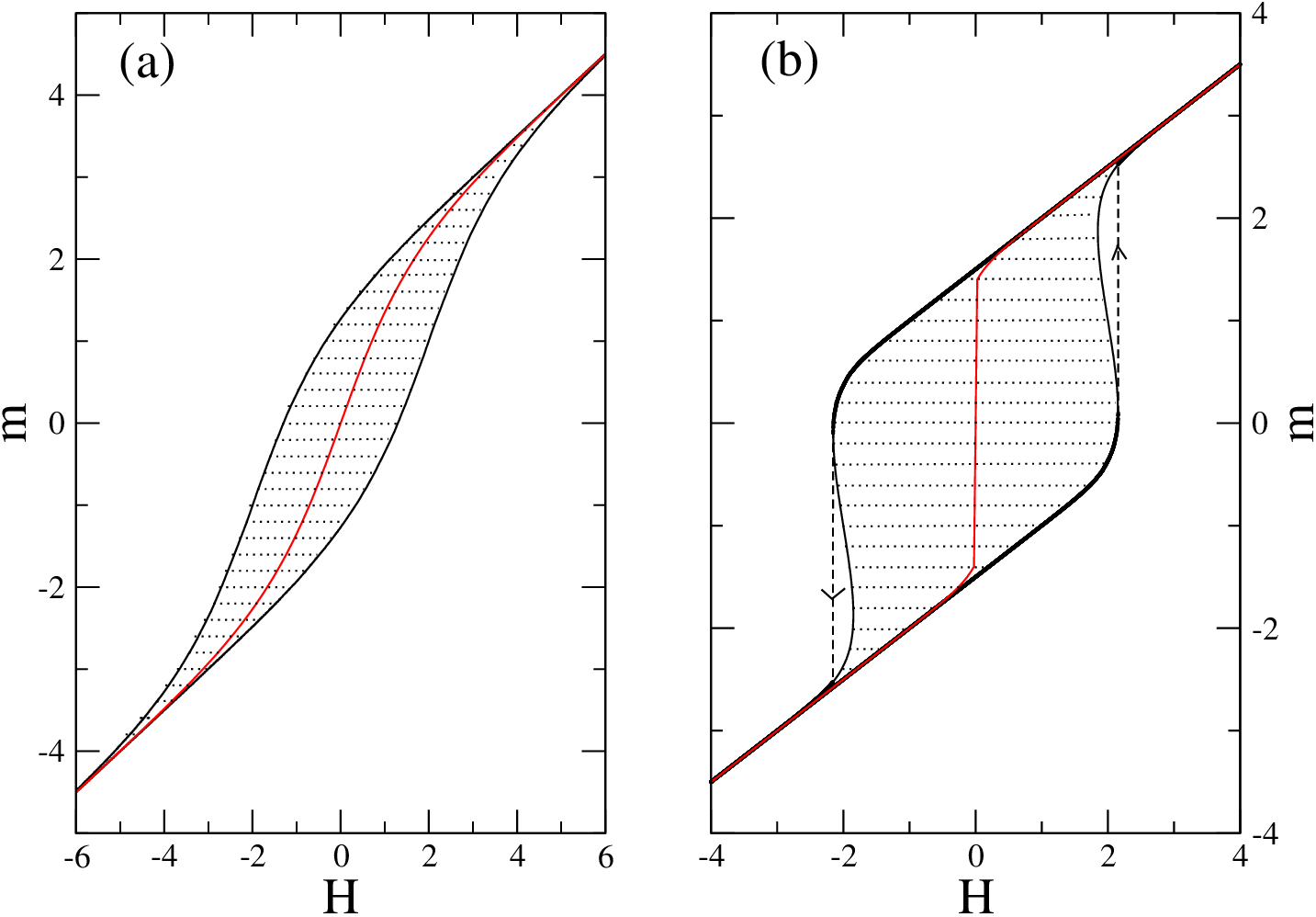}
  \caption{Mean-field hysteresis loop for the soft-spin version of the Gaussian RFIM with $k=3$ and (a) $\Delta=3$, (b) $\Delta=0.8$ ($\Delta_c\approx1.197$). In (b), the magnetization has a jump at $H=\pm H_c(0.8)\approx \pm 2.152 $. The quenched complexity $\Sigma_Q(m,H)$ is positive inside the shaded area and vanishes on the boundary. The red curve represents the ground-state magnetization. ($J$ is taken as the energy unit in all figures).}
\end{figure}

In the low-disorder regime, the jump in the ascending branch occurs at a  field $H=H_c(\Delta)$ where the slope $dm_\uparrow/dH$ diverges ($H_c(\Delta)$ is solution of the implicit equation $2kJ{\cal P}(H_c+Jm_\uparrow(H_c)-k)=k-J$).  Note that $\Delta_c$ is also the critical disorder at which a discontinuity first occurs in the equilibrium magnetization curve. The ground-state magnetization $m_{GS}(H)$ is readily obtained by noting that the lowest energy at the field $H$ is obtained when all spins with $h_i<-H-Jm$ are negative and all spins with $h_i>-H-Jm$ are positive. This yields the self-consistent equation
\begin{eqnarray}
\label{EqGS}
m_{GS}(H)= \frac{H+k}{k-J}-\frac{2k}{k-J}\int_{-\infty}^{-H-Jm_{GS}} dh {\cal P}(h) 
\end{eqnarray}
which also admits three solutions in a certain field interval for $\Delta<\Delta_c$. By definition, the ground state is the state with the lowest energy, whence the discontinuity in $m_{GS}(H)$ at $H=0$ (and the intermediate branch is unstable). 

\section{Complexity}

Whereas the computation of the quenched complexity is a difficult task when a finite number of spins interact (see e.g. \cite{DRT2005,RTP2008b}), it becomes trivial in the mean-field limit. The results, however, nicely illustrate the qualitative behavior of the complexity and its relationship to the hysteresis loop (there are also some special features of the mean-field model which will be pointed out in the following).

\subsection{Complexity as a function of field and magnetization}

We first consider the typical number of metastable states with given magnetization per spin $m$ at the field $H$, irrespective of their energy. The associated complexity $\Sigma_Q(m,H)$ is the quantity that is immediately related to the hysteresis loop, as discussed in Refs.\cite{DRT2005,PRT2008,RTP2008b}. By definition, 
\begin{align}
\label{Eq7}
\Sigma_Q(m,H)&=\lim_{N\to \infty}\frac{1}{N} \overline{\ln {\cal N}(m,H)}=\lim_{n\to 0} \frac{1}{n} \lim_{N\to \infty}\frac{1}{N}  \Big[\overline{{\cal N}(m,H)^n}-1\Big]
\end{align}
where ${\cal N}(m,H)$ is the number of metastable states with magnetization $m$ at field $H$ and the overbar denotes the average over disorder realizations (as usual, the order of the limits $N\to \infty$ and $n\to 0$ has been inverted). In the present case ${\cal N}(m,H)$ is given by
\begin{equation}
\label{Eq8}
{\cal N}(m,H)=\int \prod_{i}ds_i \prod_{i}\delta(s_i-s_i^*)\delta(\sum_i s_i-Nm)
\end{equation}
where $s_i^*$ is a solution of Eq. (\ref{Eq3}) (from a technical point of view, the simplifying feature of the double-well potential $V(s)$ described by Eq. (\ref{Eq2}) is that there is no solution of $\partial {\cal H}/\partial s_i=0$, i.e. no stationary point of the Hamiltonian that is a local maximum).
Introducing the integral representation of the second delta function in Eq. (\ref{Eq8}) and averaging over disorder gives
\begin{align}
\label{Eq9}
\overline{{\cal N}(m,H)^n}&=\frac{1}{(2i\pi)^n}\int (\prod_{a}dg_a) e^{-Nm\sum_a g_a}\prod_{i,a} \int dh_i {\cal P}(h_i)\int ds_i^a e^{g_a s_i^a} \delta(s_i^a-s_i^*) 
\end{align}
where the replica label $a$ runs from $1$ to $n$ and $g_a$ is a Lagrange multiplier associated to the constraint $\sum_i s_i^a=Nm$ in replica $a$. Using Eq. (\ref{Eq3}), we then obtain
\begin{align}
\label{Eq10}
\overline{{\cal N}(m,H)^n}= \frac{1}{(2i\pi)^n}\int \prod_{a}dg_a e^{N[\Lambda(\{g_a\})-m\sum_a g_a]}
\end{align}
where
\begin{align}
\label{Eq11}
 &\Lambda(\{g_a\})=\ln \Big \{\int dh {\cal P}(h)e^{\sum_a g_a(H+Jm+h)/k}\prod_{a} [e^{g_a} \Theta(H+Jm+h+k)+e^{-g_a} \Theta(-H-Jm-h+k)]\Big \}
\end{align}
and $\Theta(x)$ is the Heaviside function. In the large-N limit the integrals in Eq. (\ref{Eq10}) are dominated by the stationnary points $\{g_a^*\}$ that maximize $\Lambda(\{g_a\})-m\sum_a g_a$ and are solutions of 
\begin{widetext}
\begin{align}
\label{Eq12}
 m&=\frac{\partial \Lambda(\{g_a^*\})}{\partial g_a^*} \vert_m\nonumber\\
=&e^{-\Lambda(\{g_a^*\})}\int dh {\cal P}(h) e^{\sum_c g_c^*(H+Jm+h)/k}\left[\frac{H+Jm+h+k}{k}e^{g_a^*}\Theta(H+Jm+h+k)\right . \nonumber\\
&\left . +\frac{H+Jm+h-k}{k}e^{-g_a^*} \Theta(-H-Jm-h+k)\right]
\prod_{b\ne a}\left[e^{g_b^*} \Theta(H+Jm+h+k) +e^{-g_b^*} \Theta(-H-Jm-h+k)\right] \ .
\end{align}
\end{widetext}
Since all replicas are equivalent, we can set $g_a^*=g^*$ and take the limit  $n\rightarrow 0$ straight away (from now on the superscript * will be dropped in order to simplify the notations). This yields
\begin{widetext}
\begin{equation}
\label{Eq13}
 km=\int dh {\cal P}(h)\frac{(H+Jm+h+k)e^{g} \Theta(H+Jm+h+k) +(H+Jm+h-k)e^{-g} \Theta(-H-Jm-h+k)}{e^{g} \Theta(H+Jm+h+k) +e^{-g} \Theta(-H-Jm-h+k)} \ ,
\end{equation}
\end{widetext}
whence
\begin{align}
 \label{Eq14}
m&=\frac{H+k}{k-J}-\frac{2k}{k-J}\int_{-\infty}^{-H-Jm-k}dh {\cal P}(h) +\frac{k}{k-J}[\tanh(g)-1]\int_{-H-Jm-k}^{-H-Jm+k}dh {\cal P}(h) \ .
\end{align}
This finally allows us to express $g$ as a function of $m$ and $H$,
\begin{align}
\label{Eq15}
& \tanh g(m,H)=1+\frac{(k-J)m-(H+k)+2k\int_{-\infty}^{-H-Jm-k}dh {\cal P}(h)}{k\int_{-H-Jm-k}^{-H-Jm+k}dh {\cal P}(h)}\ ,
\end{align}
an equation which may have no solution for certain values of $H$ and $m$ since $|\tanh(g)|<1$. The corresponding quenched complexity is obtained from Eqs. (\ref{Eq7}) and (\ref{Eq10}) as
\begin{equation}
\label{Eq15a}
 \Sigma_Q(m,H)=\Lambda^{(1)}(g(m,H),H)-mg(m,H)
\end{equation}
 where 
\begin{align}
\label{Eq16}
 \Lambda^{(1)}(g,H)&=\lim_{n\rightarrow 0}\frac{\Lambda(g,H)}{n}\nonumber\\
&=\lim_{n\rightarrow 0}\frac{1}{n}\ln\left \{1+n\int dh {\cal P}(h)\left[g\frac{H+Jm+h}{k}+\ln \left[e^{g} \Theta(H+Jm+h+k) +e^{-g} \Theta(-H-Jm-h+k)\right]\right]+...\right\}\nonumber\\
&=g\frac{H+Jm}{k}+\int dh {\cal P}(h)\ln \left[e^{g} \Theta(H+Jm+h+k) +e^{-g} \Theta(-H-Jm-h+k)\right]\nonumber\\
&=g\frac{H+Jm+k}{k}-2g\int_{-\infty}^{-H-Jm-k}dh {\cal P}(h) + [\ln(2\cosh(g))-g]\int_{-H-Jm-k}^{-H-Jm+k}dh {\cal P}(h)   \ .
\end{align}
Using Eq. (\ref{Eq14}), we finally obtain the simple result
\begin{align}
\label{Eq17}
 \Sigma_Q(m,H)=[\ln(2\cosh(g))-g\tanh g]\int_{-H-Jm-k}^{-H-Jm+k}dh {\cal P}(h)
\end{align}
where $g(m,H)$ is solution of Eq. (\ref{Eq15}). 

\begin{figure}[hbt]
\includegraphics[width=10cm]{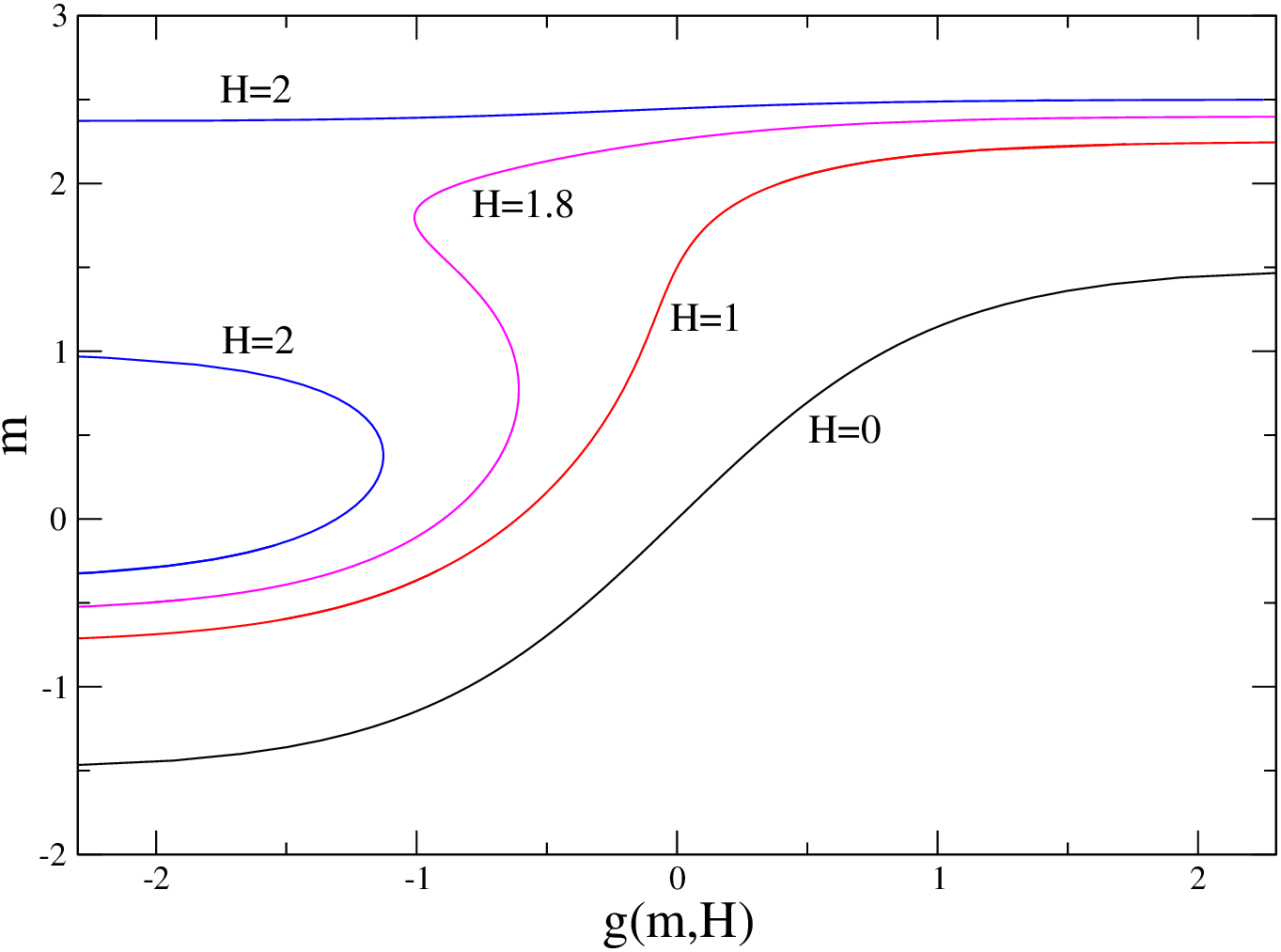}
  \caption{$g(m,H)$ as a function of $m$ for $\Delta=0.8$ and different values of $H$. For clarity, the magnetization is put on the vertical axis.}
\end{figure}

The main feature is that $\Sigma_Q(m,H)\rightarrow 0$ as $g\rightarrow \mp \infty$, and one can see immediately from Eq. (\ref{Eq15}) that Eqs. (\ref{Eq5}) and (\ref{Eq6}) that describe the ascending and descending branches of the saturation hysteresis loop, respectively, are recovered in these limits. Therefore, the complexity is not only positive in some region  {\it inside} the hysteresis loop (as a consequence of the no-passing rule\cite{note0}), but it is positive {\it everywhere} inside the loop for $\Delta>\Delta_c$ and exactly vanishes along the two branches $m_\uparrow(H)$ and $m_\downarrow(H)$. For $\Delta<\Delta_c$, this is only true before and after the jump in the magnetization. In between, the curve $\Sigma_Q(m,H)=0$ is reentrant, as shown in Fig. 1(b). This part of the curve is also described by the ``unstable'' solutions of Eqs.(\ref{Eq5}) and (\ref{Eq6}), but this is a peculiarity of the mean-field limit\cite{note1}.

The behavior of $g(m,H)$ and $\Sigma_Q(m,H)$ for $\Delta<\Delta_c$ as a function of $m$  and different values of $H$ (on the positive side) are shown in Figs. 2 and 3, respectively (in Fig. 2 we actually plot $m$ as a function of $g$ since this representation is more intuitive). One can see in Fig. 3 how the shape of $\Sigma_Q(m,H)$ is changing as $H$ increases\cite{note2}. In particular, for $H/J=2$, there are no metastable states with magnetization $1\lesssim m\lesssim 2.35$. The fact the magnetization of the metastable states cannot assume any value is at the origin of the jump in the hysteresis loop displayed in Fig. 1.

\begin{figure}[hbt]
\includegraphics[width=10cm]{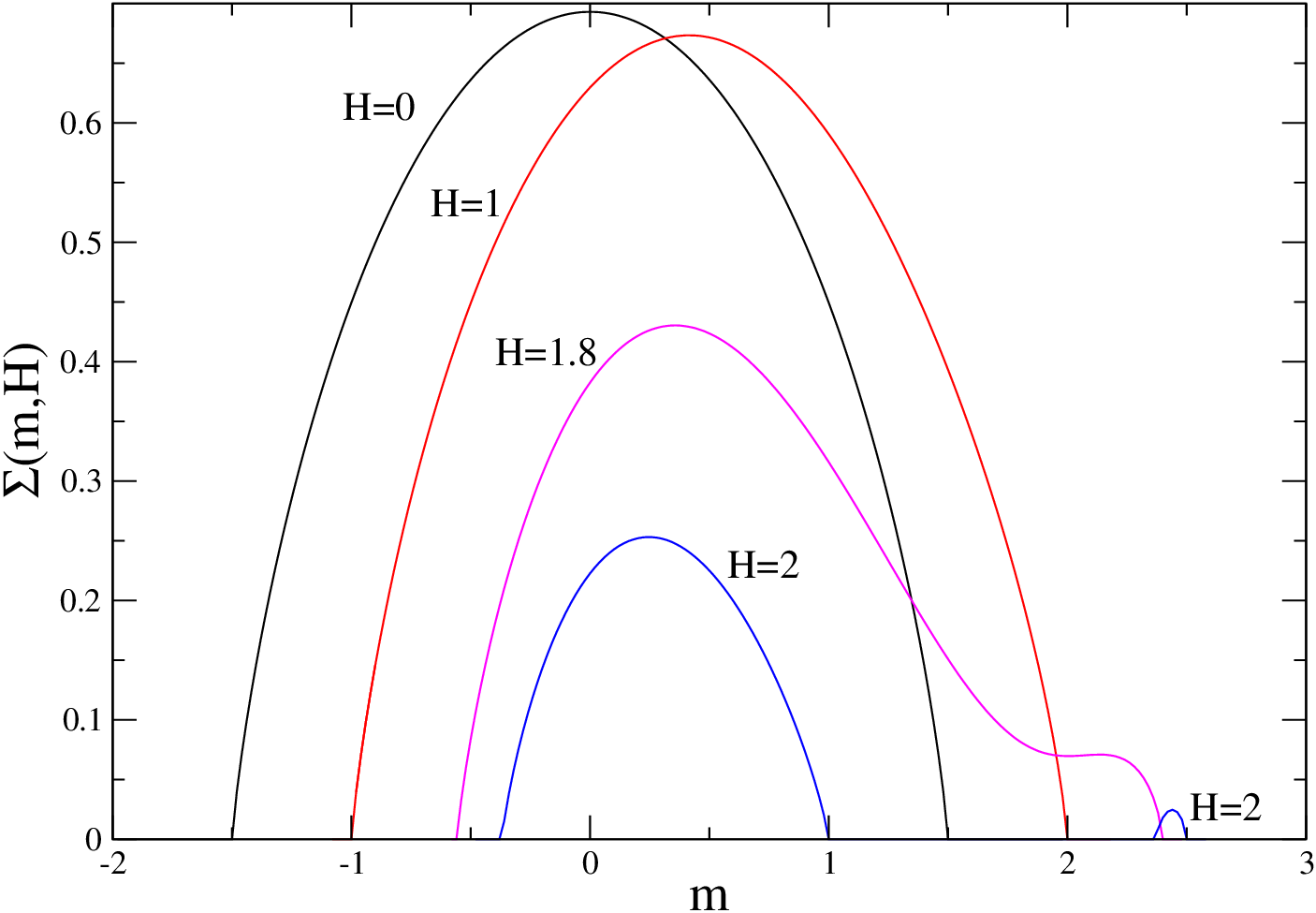}
  \caption{$\Sigma_Q(m,H)$ as a function of $m$ for $\Delta=0.8$ and different values of $H$. Note that the magnetization that corresponds to the maximum of $\Sigma_Q(m,H)$ (i.e. the typical magnetization of the metastable states at the field $H$) is not a monotonous function of $H$. For $H=2.149$, the two maxima of $\Sigma_Q(m,H)$ have the same height and the typical magnetization jumps discontinuously.}
\end{figure}

The typical (i.e. most probable) magnetization  at the field $H$ is the magnetization for which $\Sigma_Q(m,H)$ reaches its maximum. It turns out that this cannot be simply obtained by setting $g=0$ in Eqs. (\ref{Eq14}) and (\ref{Eq16}) because of the explicit dependence of $\Lambda$ on $m$ (the only exception is for $H=0$, by symmetry).  Indeed, whereas $\partial \Lambda^{(1)}(g,H)/\partial g=m$ (see Eq. (\ref{Eq12})), 
\begin{align}
\label{Eq18}
\frac{\partial \Sigma_Q(m,H)}{\partial m}&=\frac{\partial \Lambda^{(1)}(g,H)}{\partial g}\frac{\partial g}{\partial m} +\frac{\partial \Lambda^{(1)}(g,H)}{\partial m}-g -m\frac{\partial g}{\partial m}\nonumber\\
&=\frac{\partial \Lambda^{(1)}(g,H)}{\partial m}-g\nonumber\\
&\ne -g \ .
\end{align}
In other words, $\Sigma_Q(m,H)$ and  $\Lambda^{(1)}(g,H)$ are not not mutually connected by a Legendre transform, which differs from the situation in finite-connectivity models\cite{DRT2005,RTP2008b}. Eq. (\ref{Eq18}) then gives
\begin{align}
\label{Eq19}
&\frac{\partial \Sigma_Q(m,H)}{\partial m}=-g\frac{k-J}{k}+2gJ{\cal P}(H+Jm+k)+J\left[\ln(2\cosh(g))-g\right]\left[{\cal P}(H+Jm+k)-{\cal P}(H+Jm-k)\right]\ .
\end{align}
In particular,
\begin{align}
\label{Eq20}
\frac{\partial \Sigma_Q(m,H)}{\partial m}&\sim -g\left[\frac{k-J}{k}-2J{\cal P}(H+Jm\pm k)\right] \ \  \mbox{for $g\rightarrow \pm \infty$}\ ,
\end{align}
so that the curve $\partial \Sigma_Q(m,H)/\partial m=0$ joins the hysteresis loop exactly at the turning points $H=\pm H_c(\Delta)$ where the slopes of $m_\uparrow(H)$ and $m_\downarrow(H)$ diverge and the magnetization jumps. However, the actual typical magnetization is discontinuous and jumps at a field that is very close but smaller than $H_c$ in absolute value. (For the case considered in Fig. 1(b), the jump occurs at $H\approx \pm 2.149$, whereas $H_c(0.8)=\pm 2.152$. This corresponds to the situation where the two maxima of $\Sigma_Q(m,H)$ have the same height.)

\subsection{Complexity as a function of field, magnetization, and energy}

We now complete our description of the distribution of the metastable states by computing the complexity at the field $H$ for a fixed magnetization $m$ {\it and} a fixed energy per spin $e$. This is achieved by introducing an additional Lagrange multiplier $\beta$ which plays the role of the inverse temperature for the metastable configurations. Since the calculation proceeds along the same lines as before, we only quote the result,
\begin{align}
\label{Eq21}
 \Sigma_Q(m,e,H)=\int_{-H-Jm-k}^{-H-Jm+k} dh {\cal P}(h)\left\{\ln\left(2\cosh[g+\beta(H+Jm+h)]\right)-[g+\beta(H+Jm+h)]\tanh [g+\beta(H+Jm+h)]\right\}
\end{align}
where $g$ and $\beta$ are solutions of the implicit coupled equations
\begin{align}
 \label{Eq22}
m&=\frac{H+k}{k-J}-\frac{2k}{k-J}\int_{-\infty}^{-H-Jm-k}dh {\cal P}(h) +\frac{k}{k-J}\int_{-H-Jm-k}^{-H-Jm+k}dh {\cal P}(h)\left[\tanh\left[g+\beta\left(H+Jm+h\right)\right]-1\right] \ .
\end{align}
and 
\begin{align}
 \label{Eq23}
e&=\frac{Jm^2}{2}-\frac{(H+Jm)^2+\Delta^2}{2k}-(H+Jm)+2\int_{-\infty}^{-H-Jm-k}dh {\cal P}(h) (H+Jm+h)\nonumber\\
&-\int_{-H-Jm-k}^{-H-Jm+k}dh {\cal P}(h)(H+Jm+h)\left[\tanh\left[g+\beta\left(H+Jm+h\right)\right]-1\right] \  .
\end{align}
Conversely, we may consider that the complexity, the magnetization, and the energy are parametrized by $g$ and $\beta$. 
One recovers the equations of the preceding section when $\beta=0$. This corresponds to the maximum of $\Sigma_Q(m,e,H)$ at fixed $m$ and $H$. A randomly chosen metastable state with magnetization $m(g,\beta=0)$ will thus have an energy $e(g,\beta=0)$ with probability $1$ in the thermodynamic limit. 

From these equation we can now study the domain of existence of the metastable states as a function of $H$, $m$, and $e$, and determine the minimum (resp. maximum) value of the energy for fixed $H$ and $m$ below (resp. beyond) which there are no typical metastable states and $\Sigma_Q(m,e,H)=0$ (or, alternatively, the minimum and maximum values of $m$ at a given energy level $e$).  Eq. (\ref{Eq21}) tells us that  $\Sigma_Q(m,e,H)$ goes to zero when $g$ and $\beta$ go to  $\pm \infty$, but the ratio $r=g/\beta$ must be kept finite so that $m$ and $e$ can be varied continuously by controlling $r$. One finds from Eq. (\ref{Eq22}) or (\ref{Eq23}) that $|r|\le k$, and depending on whether $\beta \rightarrow +\infty$ or $ -\infty$ these equations become
\begin{align}
\label{Eq24}
m=\frac{H+k}{k-J}-\frac{2k}{k-J}\int_{-\infty}^{-H-Jm-r}dh {\cal P}(h) 
\end{align}
\begin{align}
\label{Eq25}
e&=\frac{Jm^2}{2}-\frac{(H+Jm)^2+\Delta^2}{2k}-(H+Jm)+2\int_{-\infty}^{-H-Jm-r}dh {\cal P}(h) (H+Jm+h) \ ,
\end{align}
or
\begin{align}
\label{Eq26}
m&=\frac{H+k}{k-J}-\frac{2k}{k-J}\big[\int_{-\infty}^{-H-Jm-k}dh {\cal P}(h) +\int_{-H-Jm-r}^{-H-Jm+k}dh {\cal P}(h)\big]
\end{align}
\begin{align}
\label{Eq27}
e&=\frac{Jm^2}{2}-\frac{(H+Jm)^2+\Delta^2}{2k}-(H+Jm)\nonumber\\
&+2\left[\int_{-\infty}^{-H-Jm-k}dh {\cal P}(h)(H+Jm+h) +\int_{-H-Jm-r}^{-H-Jm+k}dh {\cal P}(h)(H+Jm+h)\right] \ .
\end{align}
Both the hysteresis loop and the ground state are recovered for special values of $r$. The  ascending (resp. descending) branch of the hysteresis loop, $m_{\uparrow}(H)$ (resp. $m_{\downarrow}(H)$), is recovered from Eq. (\ref{Eq24}) for $r=-k$ (resp. $+k$) or from Eq. (\ref{Eq26}) for $r=+k$ (resp. $-k$). The equation for the ground-state magnetization, Eq. (\ref{EqGS}), is recovered from Eq. (\ref{Eq24}) for $r=0$, whereas Eq. (\ref{Eq26}) for $r=0$ gives the magnetization of the  metastable states with the maximum energy.

\begin{figure}[hbt]
\includegraphics[width=10cm]{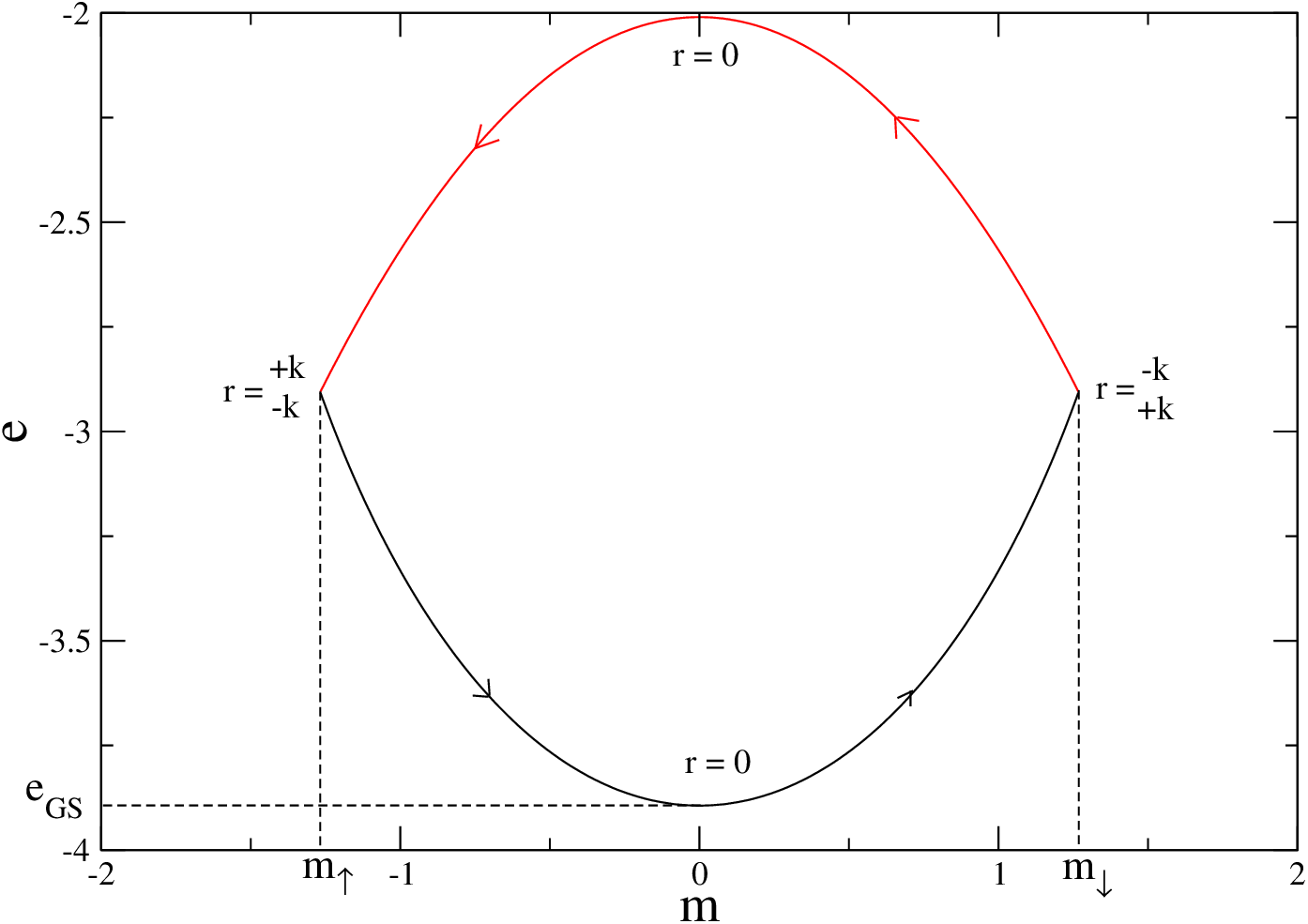}
  \caption{The curve $\Sigma_Q(m,e,H)=0$ in the magnetization-energy plane for $\Delta=3$ and $H=0$. The parts of the curve that correspond to $\beta \rightarrow +\infty$ and $\beta \rightarrow -\infty$ are drawn in black and red, respectively, and the arrows indicate the sense of variation of the parameter $r$.}
\end{figure}

In the large-disorder regime $\Delta>\Delta_c$, Eqs. (\ref{Eq24}) and (\ref{Eq26}), considered as implicit equations for $m$, have a unique solution for all values of $H$ and $r$ and the generic behavior of the curve $\Sigma_Q(m,e,H)=0$ in the plane $m-e$ is quite simple. This is illustrated in Fig. 4 for $\Delta=3$ and $H=0$ (we also show in the figure how the parameter $r$ varies along the curve).
More generally, the behavior of $\Sigma_Q(m,e,H)$ for $\Delta>\Delta_c$ is rather featureless and not worthy of special comment. Some typical results in zero external field are shown in  Figs. 7(a) and 8(a) below.

\begin{figure}[hbt]
\includegraphics[width=10cm]{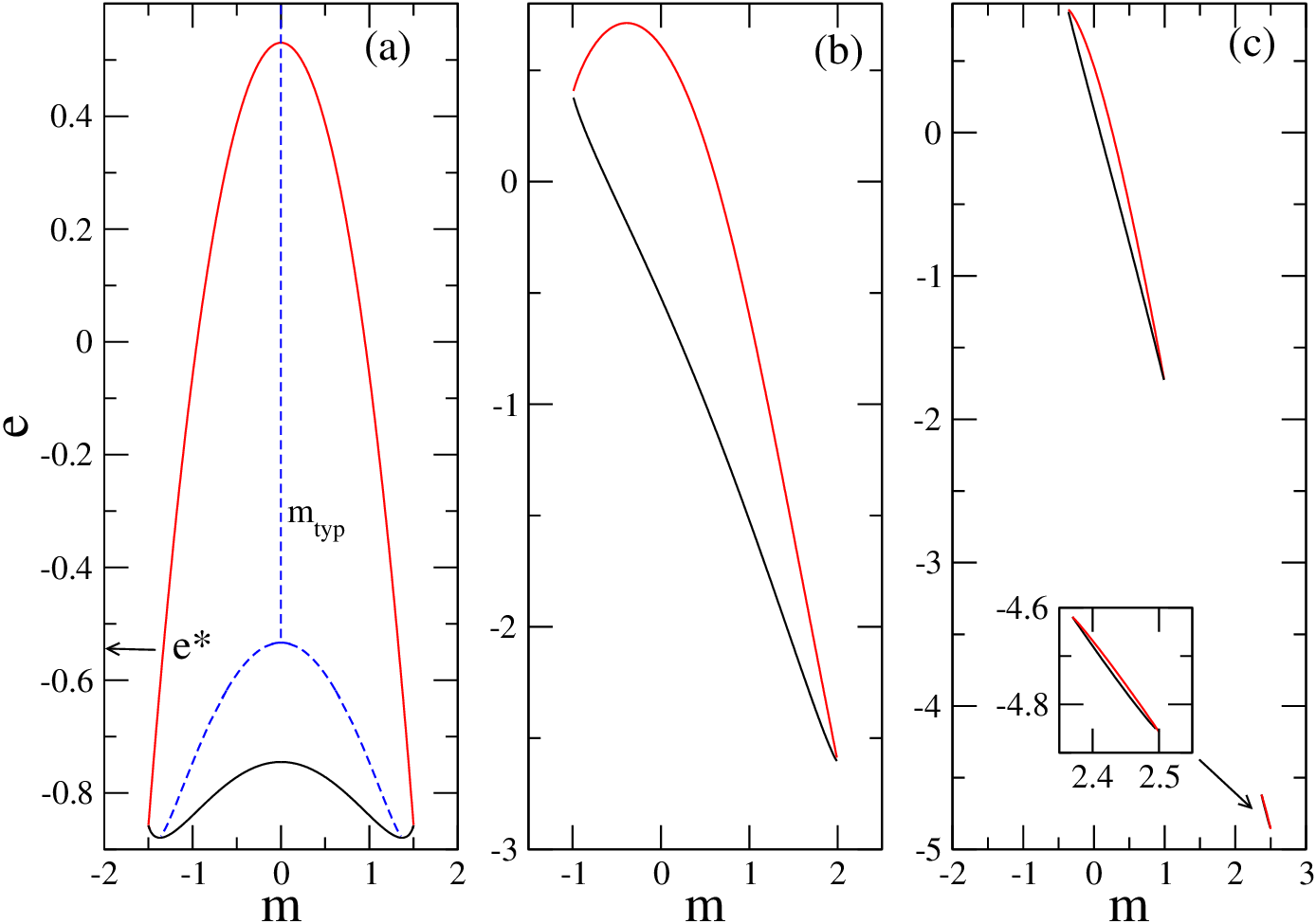}
  \caption{Same as Fig. 4 for $\Delta=0.8$ and (a) $H=0$, (b) $H=1$, (c) $H=2$. For $H=0$,  the typical magnetization of the metastable states at a given energy, i.e. the magnetization at which the complexity $\Sigma_Q(m,e,H=0)$ is maximum, is plotted as a function of energy (blue dashed line). The typical magnetization is non-zero for $e<e^*\approx-0.533$. For $H=2$,  the domain within which $\Sigma_Q(m,e,H)$ is positive is disconnected (the inset is a blow-up of the little domain on the right).}
\end{figure}

Much more interesting is the behavior in the low-disorder regime $\Delta<\Delta_c$. In this case, Eqs. (\ref{Eq24}) and (\ref{Eq26}) may have three  solutions in a certain range of $H$ and $r$ and the domain within which $\Sigma_Q(m,e,H)$ is positive is no more convex and may even break into disconnected regions, as shown in Fig. 5. This gives rise to phase transitions. 

\begin{figure}[hbt]
 \includegraphics[width=10cm]{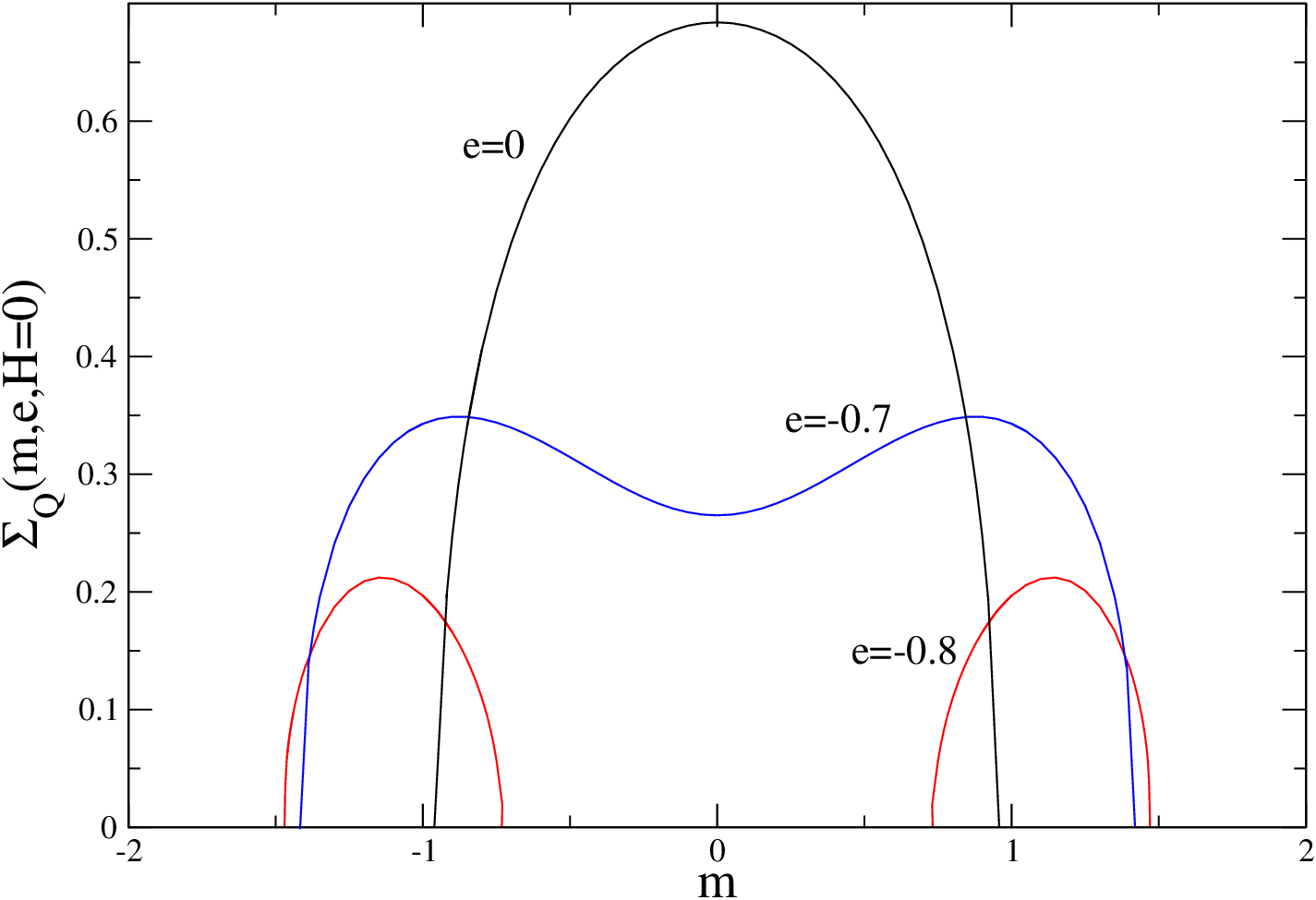}
  \caption{Complexity $\Sigma_Q(m,e,H)$ as a function of magnetization for $\Delta=0.8$, $H=0$ and three  energy values, $e=0$, $e=-0.7$, and $e=-0.8$. At low energies, the complexity has a double-peak structure and the typical metastable configurations are magnetized. Eventually, there is a range of magnetizations to which no metastable configurations can be associated.}
\end{figure}

Let us first consider the situation in zero external field.  A remarkable feature is that the typical metastable configurations have a non-zero magnetization at low energies (and hence low temperature). This is shown in Fig. 6 where we plot the complexity as a function of $m$ for $\Delta=0.8$ and three different values of the energy. While the complexity  has a single maximum at $m=0$ for $e=0$, it exhibits two maxima at symmetric values of the magnetization at lower energies. In this case, the spin configurations that dominate the distribution of the metastable states in the thermodynamic limit are  magnetized. This occurs below a certain energy $e^*(\Delta)$ (or, equivalently, for $\beta$ larger than some threshold $\beta^*(\Delta)$), as shown in  Fig. 5(a). We notice that a similar result was found numerically for the low-energy metastable configurations in the 3d RFIM\cite{BKLS2001}, which shows that this is not an artifact of the mean-field model (the same feature also occurs for the pure Ising ferromagnet on random regular graphs\cite{DL2001,BS2001}). In contrast, the typical magnetization of the metastable states is zero at all energies for $\Delta>\Delta_c$.

When $H=0$, the extrema of the complexity and the corresponding magnetizations and energies can be readily obtained by setting $g=0$ in Eqs. (\ref{Eq21})-(\ref{Eq23}) because of the symmetry $H\leftrightarrow-H$, $m\leftrightarrow -m$ (this is not true in general, as discussed previously). The maximal complexity $\Sigma_Q(e,H=0)=\max_m\Sigma_Q(m,e,H=0)$, i.e. the complexity of the metastable states irrespective of their magnetization, is plotted in Fig. 7 as a function of energy, and the corresponding typical energy is plotted in Fig. 8 as a function of the inverse temperature $\beta$  (in Figs. 7(b) and 8(b), we also show the branches that correspond to the states with zero magnetization which have a smaller complexity). 

 \begin{figure}
\includegraphics[width=10cm]{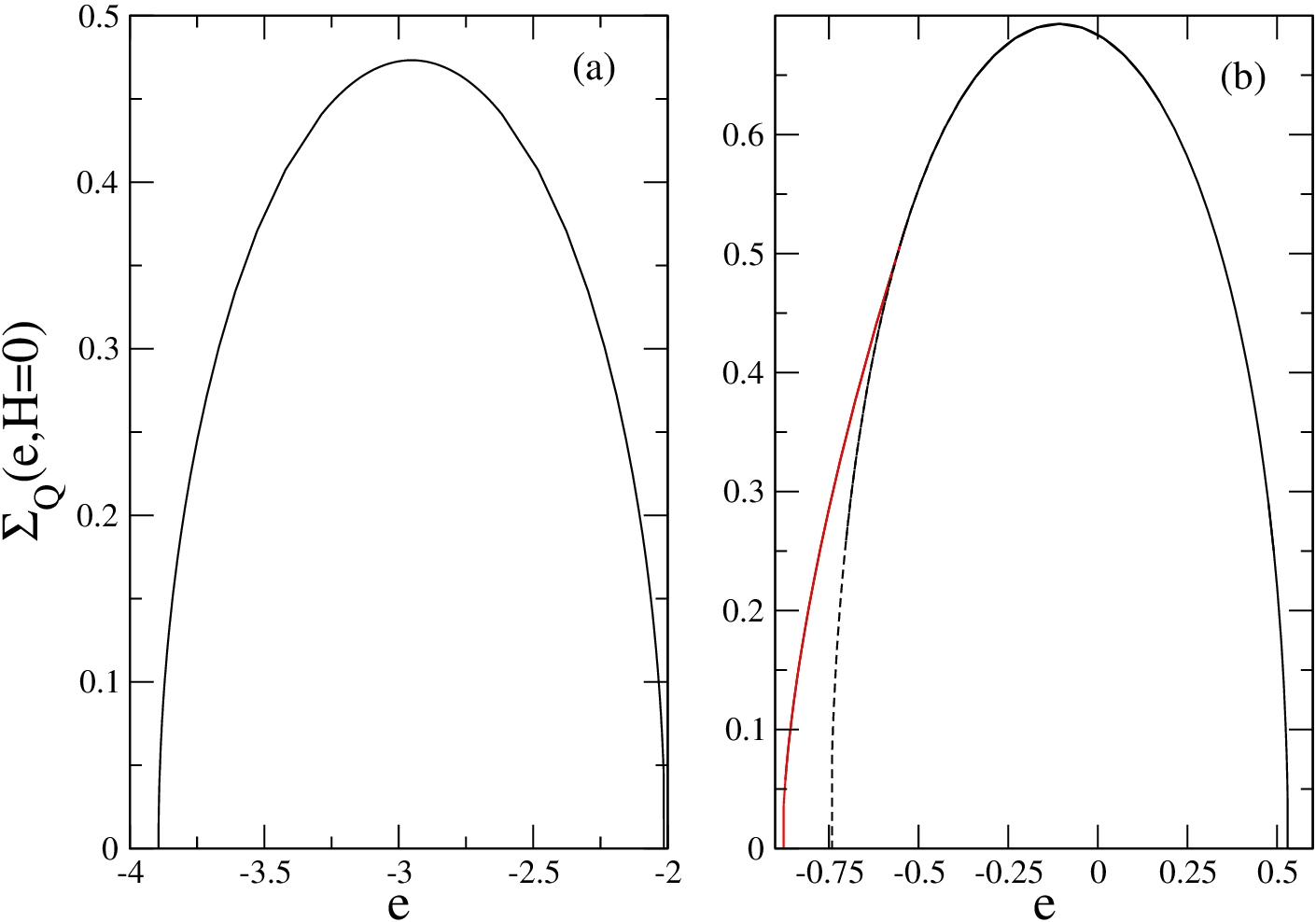}
  \caption{Complexity of the metastable states as a function of their energy $e$ (and irrespective of their magnetization) for $H=0$: (a) $\Delta=3$, (b) $\Delta=0.8$. In (b) the typical magnetization   of the metastable states is zero for $e>e^*$ and non-zero (red curve) for $e<e^*$, with $e^*\approx-0.533$. For $e<e^*$, the complexity of the metastable states with $m=0$ (dashed curve) is smaller. }
\end{figure}
\begin{figure}[hbt]
\includegraphics[width=10cm]{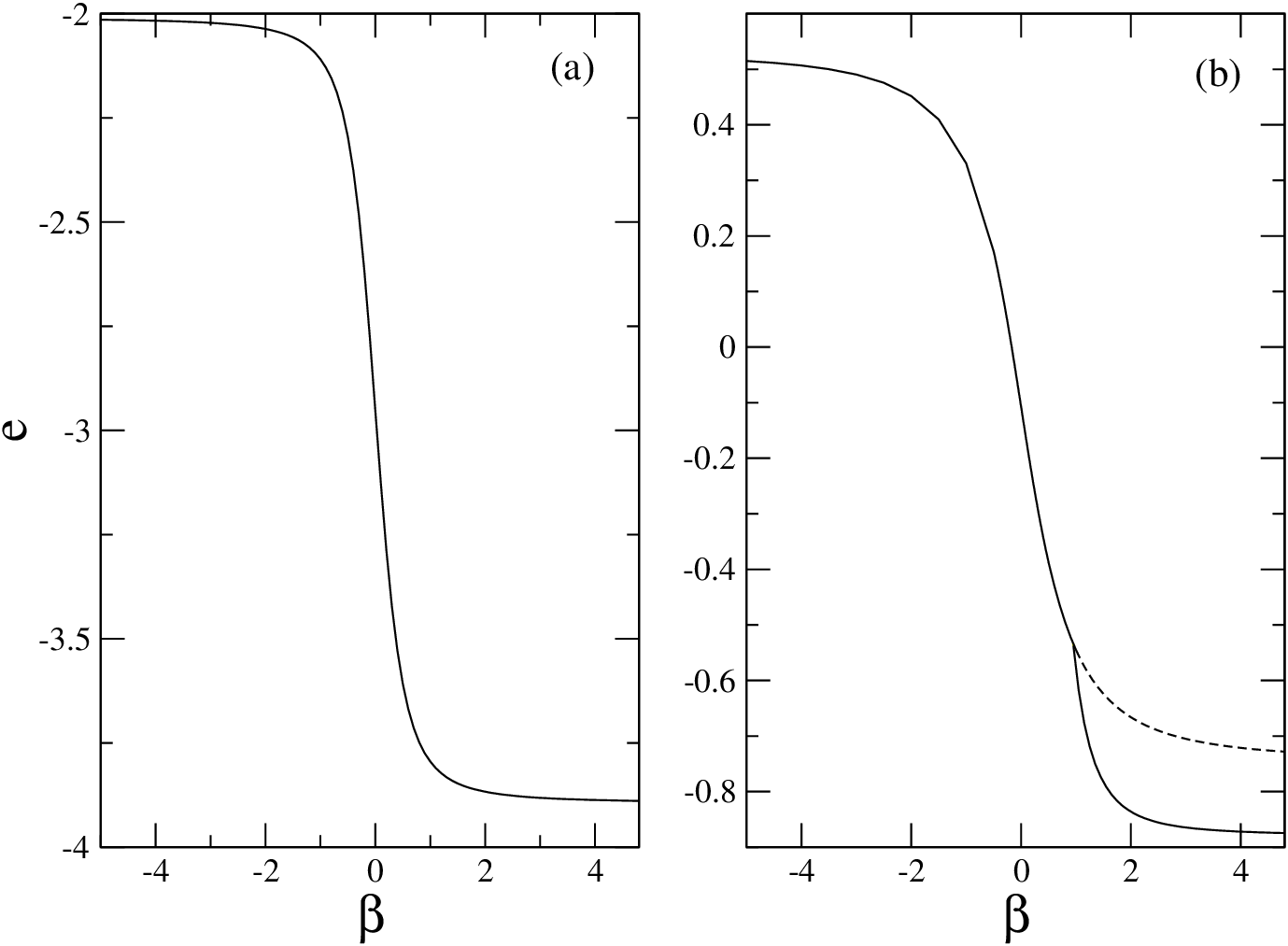}
  \caption{Typical energy of the metastable states against inverse temperature for $H=0$: (a) $\Delta=3$, (b) $\Delta=0.8$. In (b), the kink in the energy at $\beta=\beta^*\approx 0.953$ corresponds to a second-order phase transition: for $\beta>\beta^*$ the typical metastable states are magnetized (the dashed curve indicates the energy of the metastable states with $m=0$).}
\end{figure}

The transition from configurations with zero magnetization at high energies to magnetized configurations at low energies is a second-order phase transition. The inverse temperature $\beta^*(\Delta)$ is thus obtained by expanding Eq. (\ref{Eq22}) (with $H=0$ and $g=0$) around $m=0$ and cancelling the term proportional to $m$. This yields the equation
\begin{align}
\label{Eq28}
\frac{k-J}{2kJ}-{\cal P}(k)+\int_{0}^{k}dh {\cal P}'(h) \tanh(\beta^* h)=0
\end{align}
where ${\cal P}'(h)=d{\cal P}(h)/dh$. The corresponding energy $e^*(\Delta)$ is obtained by setting $H=0$, $m=0$, $g=0$, and $\beta=\beta^*(\Delta)$ in Eq. (\ref{Eq23}). Eq. (\ref{Eq28}) thus defines a second-order line in the disorder-energy and disorder-temperature planes separating a paramagnetic and a ferromagnetic phase, as shown in Figs. 9 (a) and 9(b).  This extends the equilibrium phase diagram of the model to metastable states of low energy. 

\begin{figure}[hbt]
\includegraphics[width=10cm]{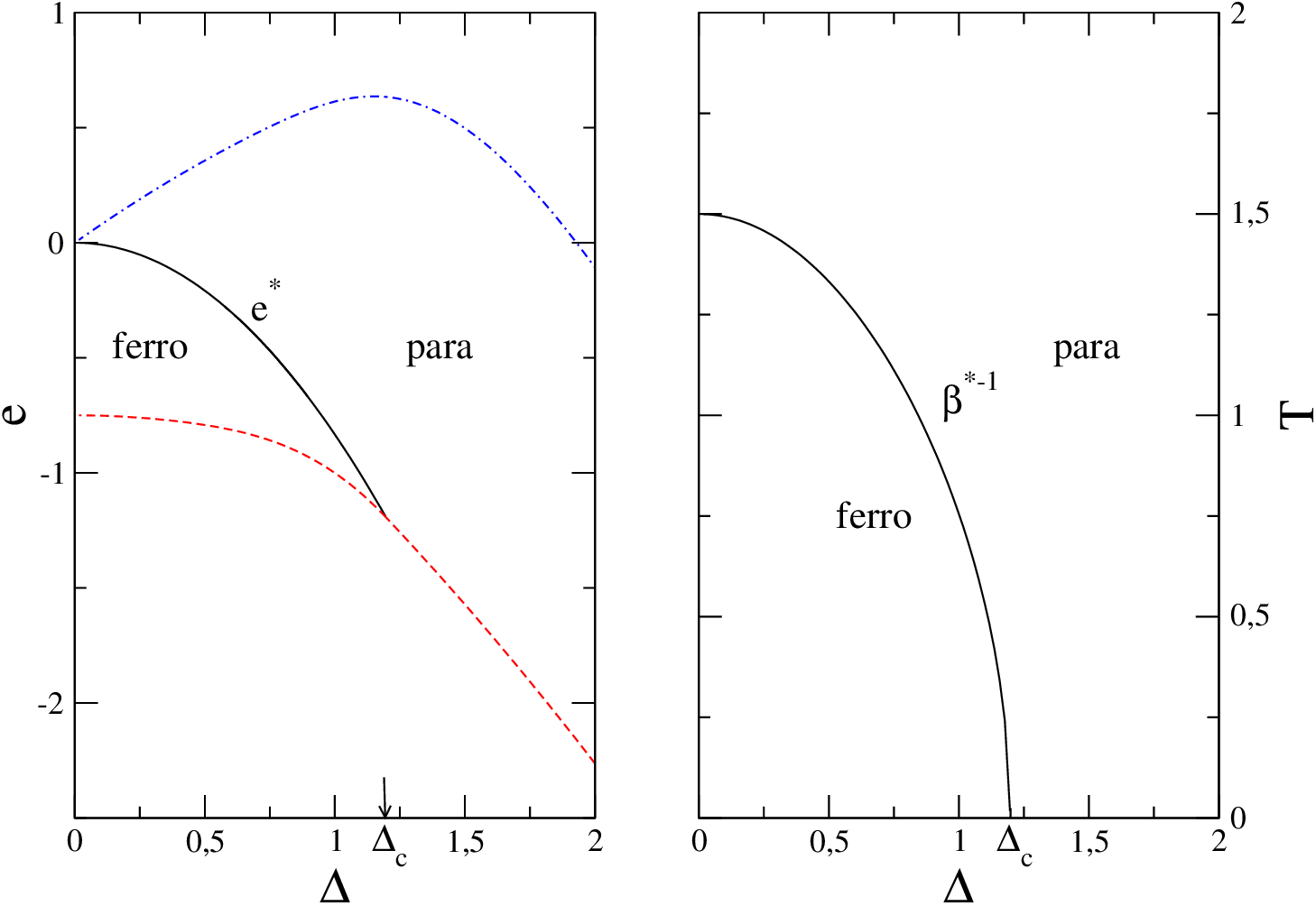}
  \caption{Phase diagram of the transition to ferromagnetic metastable states in zero external field. In (a), the energy $e^*$ at the transition is plotted against the disorder $\Delta$ (black solid line); we also show the minimum (red dashed line) and maximum (blue dotted-dashed line) allowed values of the energy. In (b), the transition ``temperature'' $1/\beta^*$ is plotted against the disorder. At low energy (or low temperature) and low disorder, the typical metastable states are magnetized. Otherwise, their typical magnetization is zero.}
\end{figure}

In non-zero field, the most salient feature is that the domain within which $\Sigma_Q(m,e,H)$ is positive may be disconnected, as shown in Fig. 5(c). This is of course in agreement with the  results of the preceding section, and, as already emphasized, is at the origin of the jump in the hysteresis loop.

\section{H-driven and M-driven protocols}

Another issue that is worth discussing in the present context concerns the influence of the driving mode on the nonequilibrium response of the system. This issue has been discussed in detail in the case of sandpiles or driven interfaces in disordered media\cite{DMVZ2000} (for instance when an elastic chain is driven adiabatically at constant force or constant velocity on a disorder substrate\cite{LZH2001}). Experimentally, when the variable conjugated to the force (or field or stress) is externally controlled instead of the force itself, one observes a reentrant hysteresis loop as well as large fluctuations of the induced force\cite{B1998,BRIMPV2007}. The magnetization-driven RFIM exhibits this  behavior in the low-disorder regime\cite{IRV2006}, and it has been suggested that the resulting hysteresis curve follows the  boundary of the domain of existence of the typical metastable states in the thermodynamic limit. We can now test this assumption in the mean-field model since we have an exact analytical description of the curve $\Sigma_Q(m,H)=0$. Note that the problem of finding the magnetic field that produces a desired magnetization function is known in the hysteresis literature as the ``inverse Preisach problem''\cite{BBSV1998}.

 In the present case, the magnetization-driven (or $M$-driven) protocol is defined as follows. As the magnetization $m=M/N$ is slowly varied, the system attempts to minimize (at least locally) its internal energy ${\cal U}(\{s_i\})=-(J/2N)\sum_{i\ne j}s_is_j-\sum_i h_is_i+\sum_iV(s_i)$ (there is no externally imposed magnetic field). Each spin $s_i$ thus satisfies $\partial {\cal U}/\partial s_i=0$ with the global constraint $\sum_{i=1}^N s_i=Nm$. This simple problem of constraint optimization is readily solved by introducing the quantity 
\begin{equation}
\label{Eq50}
{\cal L}={\cal U}-\lambda \left\{\sum_{i=1}^N s_i-Nm\right \}
\end{equation}
where $\lambda$ is a Lagrange multiplier that has the meaning of a magnetic field coupled to the extensive variable $\sum_{i=1}^N s_i$. Minimizing ${\cal U}(\{s_i\})$ with the constraint on the magnetization amounts to solve simultaneously the $N + 1$ coupled equations $\partial{\cal L}/\partial s_i = 0$ and $\partial {\cal L}/\partial \lambda=0$, 
\begin{align}
\label{Eq51}
s_i(\lambda)- \mbox{sign}(s_i(\lambda))= \frac{\lambda+Jm+h_i}{k} 
\end{align}
with 
\begin{equation}
\label{Eq52}
\sum_{i=1}^N s_i(\lambda)=Nm \ .
\end{equation}
Comparing to Eq. (\ref{Eq3}), it is clear that the spin configurations $\{s_i^*\}$, solutions of Eq. (\ref{Eq51}), are mimima of the  Hamiltonian ${\cal H}={\cal U}-H\sum_is_i$ for the special value of the field $H=\lambda^*(m)$ that satisfies the constraint equation, Eq. (\ref{Eq52}). The Lagrange multiplier $\lambda$ can be eliminated from Eq. (\ref{Eq51}) by summing over $i$, which gives
\begin{equation}
\label{Eq53}
\lambda=(k-J)m-\frac{k}{N}\sum_{i=1}^N \mbox{sign}(s_i)\ ,
\end{equation}
neglecting the contribution $(1/N) \sum_i h_i$ that vanishes in the thermodynamic limit. Replacing in Eq. (\ref{Eq51}) yields
\begin{align}
\label{Eq54}
s_i -\mbox{sign}(s_i)=m+\frac{h_i}{k}-\frac{1}{N}\sum_{j=1}^N \mbox{sign}(s_j)\ .
\end{align}
(Note that this equation can be simply rewritten as $\sigma_i =\mbox{sign}(f_i)$ where $f_i=\sigma_i+m+h_i/k-(1/N)\sum\sigma_j$ and $\sigma_i$ is the auxiliary Ising variable defined by $\sigma_i=\mbox{sign}(s_i)$.) The crucial feature in Eq. (\ref{Eq54}) is the presence of the {\it antiferromagnetic} contribution $-(1/N)\sum_{j=1}^N \mbox{sign}(s_j)$ which plays the role of an infinite-range demagnetizing field (see Ref.\cite{PTZ2008} for a more detailed discussion of this feature in a slightly different framework). As is well known, such a term may have a dramatic influence on the behavior of magnetic systems\cite{B1998} and models of interface growth in disordered media show that it may result in self-organized criticality\cite{KS2000,DMVZ2000,UMM1995,ZCDS1998}.

There may be of course many solutions to Eq. (\ref{Eq54}) (which simply characterizes all metastable states with magnetization $m$) and one needs to define a relaxation dynamics to go from a metastable configuration at $m$ to the ``nearest'' one at $m+\delta m$. The most natural algorithm consists in searching for the fixed point of the iterative map 
\begin{align}
\label{Eq55}
s_i^{(n+1)}-\mbox{sign}(s_i^{(n+1)})= m+\delta m+h_i/k-(1/N)\sum_j \mbox{sign}(s_j^{(n)})
\end{align}
applied to all $s_i$ in parallel, where $s_i^{(0)}=s_i^*(m)$ is the converged value at magnetization $m$. Since this equation has no negative solution when the right-hand-side is larger than $1$ and no positive solution when it is smaller than $-1$, the evolution of the spins (and, as a consequence, of the induced field) is discontinuous. Remarkably, it turns out that  convergence to the fixed point is reached after only one or two iterations. The first iteration gives
\begin{align}
\label{Eq56}
s_i^{(1)}=s_i^{(0)}+\delta m+[\mbox{sign}(s_i^{(0)}+\delta m)-\mbox{sign}(s_i^{(0)})]
\end{align}
which implies, e.g. when increasing $m$ from a large negative value, that the spins in the range $-\delta m< s_i^{(0)}<0$ become positive. However, these spins never become negative again (i.e. there are no ``back-flips''). The second iteration then yields
\begin{align}
\label{Eq57}
s_i^{(2)}=s_i^{(1)}-(1/N)\sum_j [\mbox{sign}(s_j^{(1)})-\mbox{sign}(s_j^{(0)})]
\end{align}
and there is no need for a third iteration, i.e. $s_i^*(m+\delta m)=s_i^{(2)}$. This implies from Eq. (\ref{Eq53}) that the induced field at $m+\delta m$ along the ascending branch is simply given by 
\begin{align}
\label{Eq58}
H_{\uparrow}(m+\delta m)=H_{\uparrow}(m)+(k-J)\delta m-(k/N)\sum_i [\mbox{sign}(s_i^*(m)+\delta m)-\mbox{sign}(s_i^*(m))]\ .
\end{align}

 \begin{figure}[hbt]
\includegraphics[width=10cm]{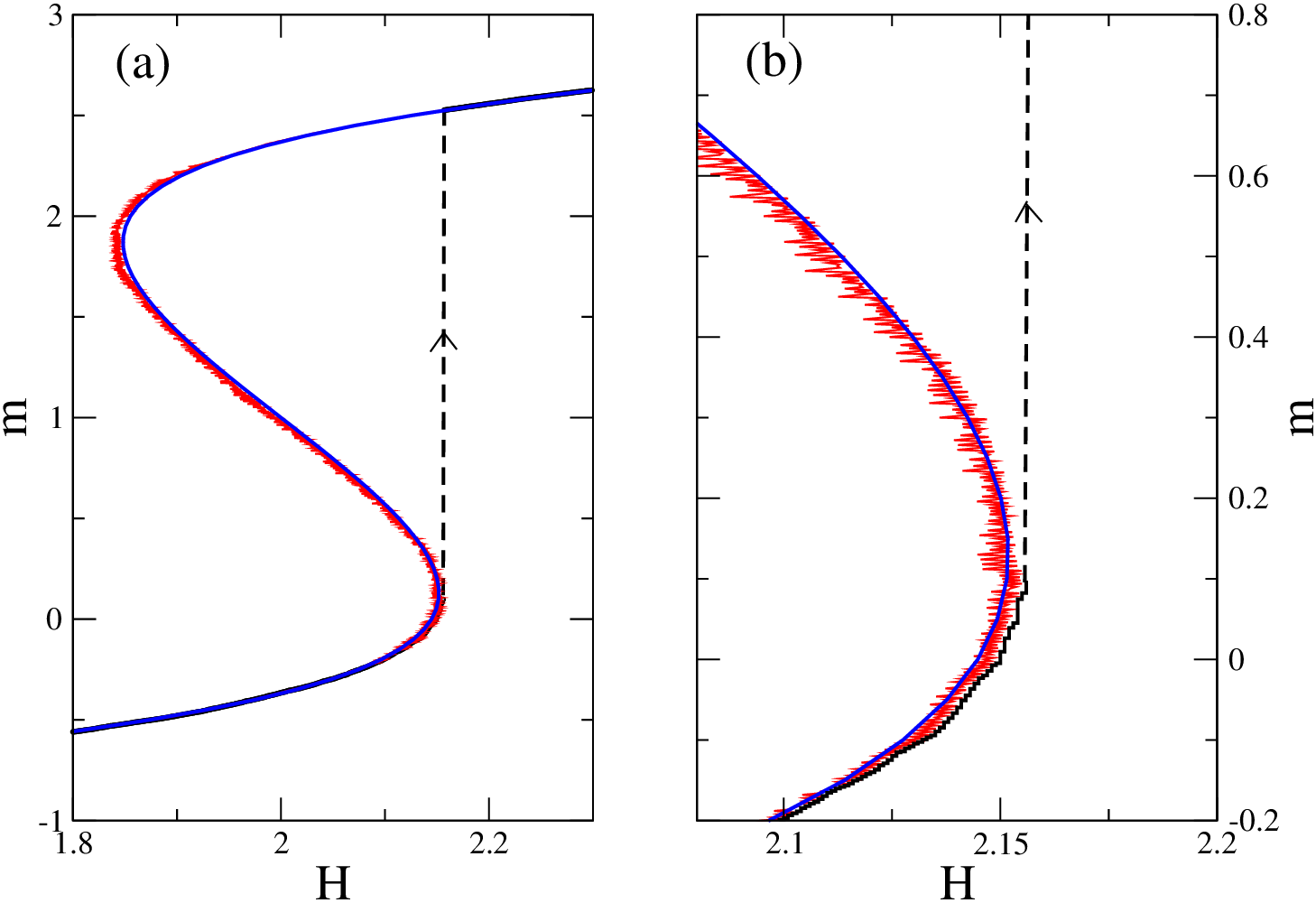}
  \caption{Comparison of the nonequilibrium responses obtained by increasing adiabatically the field $H$ (black line) or the magnetization $m$ (red line) from $-\infty$ to $+\infty$ in a single disorder sample of size $N=50000$ for $\Delta=0.8$. For clarity, the magnetization is on the vertical axis in both cases. The step sizes were $\delta H=0.001$ and $\delta m=0.002$, respectively.  The $H$-driven response has a jump whereas the $M$-driven response is reentrant (the figure (b) is a blow-up in the vicinity of the knee). The blue line represents the curve $\Sigma_Q(m,H)=0$, solution of Eq. (\ref{Eq5}) (the discrepancy with the $H$-driven trajectory before the jump is due to finite-size effects).}
\end{figure}

A typical example of the nonequilibrium response obtained with this protocol in a single disorder sample in the low-disorder regime is shown in Fig. 10 and compared to the result of the $H$-driven protocol\cite{note3}.  One can see that the $M$-driven response is reentrant whereas the $H$-driven response displays a jump at $H\approx 2.15$, in accordance with the infinite-$N$ behavior described in section 2 (there are also many small avalanches that can be better seen in Fig. 10(b)).  Moreover, the induced field $H(m)$ follows very closely the curve $\Sigma_Q(m,H)=0$ that defines the boundary of the domain of existence of the metastable states in the thermodynamic limit. They are fluctuations in $H(m)$ but the $M$-driven trajectory is always {\it inside} the $H$-driven hysteresis loop, as can be seen in Fig. 4(b). Indeed, for in given disorder realization, there are no metastable states outside the hysteresis loop\cite{note0}. 

The fluctuations in $H(m)$ decrease with $N$ and we can actually derive from Eq. (\ref{Eq58}) the exact equation of the $M$-driven trajectory in the limit $N\rightarrow \infty$, $\delta m\rightarrow 0$. Indeed, $(1/2N)\sum_i [\mbox{sign}(s_i^*(m)+\delta m)-\mbox{sign}(s_i^*(m))]$ is just the fraction of spins that are negative at magnetization $m$ and that become positive at $m+\delta m$. In the thermodynamic limit, the corresponding probability for the random field $h$ is
\begin{align}
\label{Eq59}
\int_{-H_{\uparrow}(m+\delta m)-J(m+\delta m)+k}^{-H_{\uparrow}(m)-Jm+k}dh {\cal P}(h)\sim [J+\frac{dH_{\uparrow}(m)}{dm}]{\cal P}(H_{\uparrow}(m)+Jm-k)\ \delta m \  \ \mbox{for}\ \delta m \rightarrow 0 \ ,
\end{align}
and inserting this result in Eq. (\ref{Eq58}) we obtain the differential equation 
\begin{align}
\label{Eq60}
\frac{dH_{\uparrow}(m)}{dm}=\frac{k-J -2kJ{\cal P}(H_{\uparrow}(m)+Jm-k)}{1+2k{\cal P}(H_{\uparrow}(m)+Jm-k)} \ .
\end{align}
It is easy to see that the solution of this equation is given by  Eq. (\ref{Eq5}), considered as an implicit equation for the magnetic field as a function of the magnetization, i.e. $H_{\uparrow}(m_{\uparrow}(H))=H$. As shown previously, this equation defines the curve $\Sigma_Q(m,H)=0$ in the $H-m$ plane.  Therefore, the  $M$-driven hysteresis loop exactly follows the boundary of the domain of existence of the typical metastable states in the thermodynamic limit, including along  the reentrant part in the low-disorder regime, in agreement with the conjecture of Ref.\cite{IRV2006}. Moreover,  the $H$-driven and $M$-driven loops coincide in the large-disorder regime.

\section{Summary and conclusions}

In this work we have studied the soft-spin version of the random-field Ising model in the mean-field limit and computed analytically the quenched complexity  $\Sigma_Q(m,e,H)$ of the metastable states as a function of their magnetization and energy in the presence of an external magnetic field.  We have especially focused on the domain ${\cal D}(m,e,H)$ within which the complexity is positive and the number of metastable states grows exponentially with the system size (and the domain ${\cal D}(m,H)$ associated to $\Sigma_Q(m,H)=\max_e\Sigma_Q(m,e,H)\ge 0$) . The results can be summarized as follows:

1) At large disorder, the domain ${\cal D}(m,e,H)$  is convex. The quenched complexity  $\Sigma_Q(m,H)$ is then strictly positive everywhere inside  the   saturation hysteresis loop and  it vanishes along the loop. Moreover, the system responds identically when driven by the magnetic field ($H$-driven protocol) or the magnetization ($M$-driven protocol).

2) At low disorder, the domain ${\cal D}(m,e,H)$  is non-convex and disconnected. The curve $\Sigma_Q(m,H)=0$ has a reentrant part and coincides with the  $M$-driven hysteresis loop.

3) In zero external field, there is a second-order line in the disorder-energy plane that separates a paramagnetic and a ferromagnetic phase: the ground state and the low-lying metastable states are magnetized at low disorder.

This study provides a pedagogical illustration of the fact that the zero-temperature, nonequilibrium response of the system to a slow external driving can be defined unambiguously from a statistical description of the metastable states in some appropriate limit.
This suggests that  the  ground state and  the metastable states along the hysteresis loop can be studied within a unique and purely static theoretical framework, which may offer a possible new starting point to investigate the relationship between the critical behavior of the RFIM in and out of equilibrium\cite{note5}.

To which extent the mean-field picture remains valid when the interaction is short-ranged ?  It is of course impossible in this case to perform a detailed study of the distribution  of the metastable states in energy or magnetization.  However, previous numerical\cite{RTP2008a} and analytical\cite{DRT2005,RTP2008b} studies suggest that the essential qualitative features observed in the  mean-field model survive in finite dimension, in particular the presence of a gap in the  magnetization of the metastable states at low disorder, gap which is  at the origin of  the jump along the field-driven hysteresis loop (note that in finite connectivity models, this gap is not related to the existence of  an ``unstable''  solution to the self-consistent equations describing the hysteresis loop\cite{note1}: this solution is an artifact of the mean-field limit).
There is also numerical evidence that the low-energy metastable states of the $3$d RFIM in zero field are magnetized  at low disorder\cite{BKLS2001}, suggesting that the phase diagram of the metastable states as a function of energy and disorder is qualitatively similar to the one displayed in Fig. 9(a). 

On the other hand, there are also some noticeable new features in finite-connectivity models. Firstly, the critical values of the disorder associated to the equilibrium (ground-state) and nonequilibrium (hysteresis) phase transitions become distinct\cite{CADMZ2004}, with $\Delta_c^{hyst}<\Delta_c^{GS}$. Secondly, in the low-disorder regime, the complexity may not be zero along the whole boundary of the domain ${\cal D}(m,H)$ (this is still putative but supported by analytical calculations on the Bethe lattice at the order $1/z$)\cite{RTP2008b}. Thirdly, the geometry of the lattice may change qualitatively the nonequilibrium response of the system obtained by controlling the magnetization.  Preliminary simulations on a cubic lattice show a feature that was noticed in previous work\cite{IRV2006} and that has been recently discussed  in Ref.\cite{PTZ2008}: when the disorder is low enough, the induced field $H(m)$ exhibit strong fluctuations around a certain value $H_d$ that does not depend on the magnetization.  As a result, the domain ${\cal D}(m,H)$  has a very peculiar shape in this regime. The behavior around  $H_d$ actually corresponds to interface configurations in real space, and the fluctuations are similar to those observed at a critical depinning transition.  This is an interesting  issue which is worth investigating further.

\section{Acknowledgements}

M.L.R. wish to thank F. J.  P\'erez-Reche, G. Tarjus, and E. Vives for useful conversations and comments. We are grateful to G. Tarjus for a careful reading of the manuscript.


\begin{thebibliography}{10}
\bibitem{note3} Note that the metastable state that is reached when going from $m$ to $m+\delta m$ depends on the value of the step $\delta m$, a problem that does no occur when solving iteratively Eq. (\ref{Eq3}) ($H$-driven protocol). This comes from the fact that the $M$-driven algorithm does not satisfy  the ``no-passing'' rule\cite{M1992,S1993} because of the presence of the antiferromagnetic contribution $-(1/N)\sum_{j=1}^N \mbox{sign}(s_j)$.
\end{thebibliography}
\end{document}